# Universal Gelation of Metal Oxide Nanocrystals via Depletion Attractions

Camila A. Saez Cabezas[1], Zachary M. Sherman[1], Michael P. Howard[1], Manuel N. Dominguez[1,2], Shin Hum Cho[1], Gary K. Ong[1], Allison Green[1], Thomas M. Truskett[1,3,*], and Delia J. Milliron[1,*]

[1] McKetta Department of Chemical Engineering, University of Texas at Austin, Austin, Texas, 78712-1589
[2] Department of Chemistry, University of Texas at Austin, Austin, Texas, 78712-1589
[3] Department of Physics, University of Texas at Austin, Austin, Texas, 78712-1589

*Corresponding Authors

Email: truskett@che.utexas.edu, milliron@che.utexas.edu

Abstract

Nanocrystal gelation provides a powerful framework to translate nanoscale properties into bulk materials and to engineer emergent properties through the assembled microstructure. However, many established gelation strategies rely on chemical reactions and specific interactions, e.g., stabilizing ligands or ions on the nanocrystals' surfaces, and are therefore not easily transferrable. Here, we report a general gelation strategy via non-specific and purely entropic depletion attractions applied to three types of metal oxide nanocrystals. The gelation thresholds of two compositionally distinct spherical nanocrystals agree quantitatively, demonstrating the adaptability of the approach for different chemistries. Consistent with theoretical phase behavior predictions, nanocrystal cubes form gels at a lower polymer concentration than nanocrystal spheres, allowing shape to serve as a handle to control gelation. These results suggest that the fundamental underpinnings of depletion-driven assembly, traditionally associated with larger colloidal particles, are also applicable at the nanoscale.

Colloidal nanocrystals are functional building blocks that exhibit remarkable properties inherent to the nanoscale[1-5] and can be used to self-assemble structures over multiple length scales.[6] Considering the influence of the spatial arrangement of neighboring nanocrystals on nanoscale optoelectronic properties (e.g., localized surface plasmon resonance,[7-9] photoluminescence,[10-12] and photocatalysis[13-16]), self-assembly methods have become powerful tools to diversify and even enhance properties in macroscopic materials and devices.[6] Among these methods, nanocrystal gelation assembles self-supported, open, and percolated networks and has gained significant interest in the past decade owing to its exceptional ability to translate and retain nanoscale properties in a bulk material compared to denser assemblies.[17-19] Nanocrystal gelation also offers structural control over multiple length scales. Both the local structure around an individual nanocrystal (e.g. volume fraction, valence, and interparticle distance) and the global topology of the network may be tuned by changing the physicochemical interactions of the building blocks.[20-23] Therefore, this framework presents opportunities for elucidating and designing structure-property relationships.

Nanocrystal gelation involves the controlled aggregation of colloidal dispersions and has been conventionally achieved by either removing the stabilizing ligands from the colloid surfaces or creating chemical bridges between the surface-bound ligands.[17,19,24-26] Although controlled ligand removal has been adapted for gelation of different types of nanocrystals and ligand combinations, precise structural control is frustrated by fast aggregation kinetics that typically lead to sedimented gels[17,19] or sintering into continuous wire-like networks.[27-29] Inducing controlled gelation using chemical linking is a viable strategy to circumvent these challenges, but this approach is limited to specific surface chemistries that are not easily adaptable across noble metals, metal chalcogenides, and metal oxides without the development of customized ligand exchange strategies and ligand-linker combinations. Recently, efforts to generalize nanocrystal gelation have leveraged electrostatic interactions to mediate the assembly of different single-component and composite gels.[30,31] However, non-specific physical interactions remain largely unexplored in nanocrystal gel systems despite their potential to facilitate assembly of different nanocrystal sizes, complex shapes, hybrid nanocrystals, and combinations thereof.

Depletion attractions are physical and purely entropic forces that offer a general tunable strategy to assemble colloidal gels that does not depend on the composition and surface chemistry of the building blocks. In principle, the strength and range of attraction can be easily and independently controlled by changing the relative sizes of the primary colloid and the smaller cosolute (depletant) and the depletant concentration. Experimental soft material systems such as polymer microspheres[32-36] and surfactant-stabilized emulsions[37] have leveraged the intrinsic versatility of depletion attractions to form gels with different colloids in the presence of the same depletant and vice versa, but this capability has not been demonstrated with inorganic materials, especially at the nanoscale. Moreover, the strategic selection of particle shape, which dictates the geometry of densest packing and therefore determines the magnitude of the overlap excluded volume, allows further control of the strength of attractions and imparts directionality to the interaction.[38-40] In particular, this shape-dependence has been used to selectively aggregate faceted particles in mixed dispersions and separate them from spherical particles.[38,39] While depletion-mediated superlattice assembly of microcubes,[41] nanoprisms,[42] and nanopolyhedra[43] has been studied, to our knowledge, gelation with faceted particles has not been reported before.

In this work, we demonstrate a universal approach to nanocrystal gelation using depletion attractions. We expand on a protocol that we developed previously[44] based on the combination of long-range electrostatic repulsions and polymer-mediated short-range depletion attractions. As a model system, we use metal oxide nanocrystals of similar size, but different compositions and shapes: iron oxide ($FeO_x$) spheres,[45] tin-doped indium oxide ($Sn:In_2O_3$) spheres,[46] and fluorine, tin-codoped indium oxide ($F,Sn:In_2O_3$) cubes.[47] $FeO_x$, $Sn:In_2O_3$, and $F,Sn:In_2O_3$ gels were assembled by adding the same polyethylene glycol (PEG) depletant to charge-stabilized nanocrystal dispersions in acetonitrile. We assess our proposed gelation mechanism via PEG-induced depletion attractions and rationalize the phase behavior observed in all three systems with theoretical predictions of the phase behavior of the nanocrystal-depletant mixtures. The gelation thresholds for $Sn:In_2O_3$ and $FeO_x$ spheres are in quantitative agreement and do not depend on the nanocrystal composition. Consistent with calculated spinodal boundaries for spherical and cubic colloids, $F,Sn:In_2O_3$ nanocubes form a gel at a lower PEG concentration than $Sn:In_2O_3$ nanospheres. Small-angle X-ray scattering (SAXS) confirms that the gels form percolated networks that scatter as mass fractals. Samples prepared with lower PEG concentrations remain flowing dispersions and appear by SAXS as dispersed nanocrystals with minimal clustering, as predicted theoretically.

FeO$_x$, Sn:In$_2$O$_3$, and F,Sn:In$_2$O$_3$ nanocrystals of uniform size and shape (Figure 1a-c) were synthesized using established colloidal methods. The native, hydrophobic oleate ligands on the surface of the metal oxides were chemically removed using tetrafluoroborate salts[48,49] to produce charge-stabilized nanocrystal dispersions in polar solvents. A detailed description of experimental methods and characterization techniques is included in the Supporting Information. Fourier transform infrared spectroscopy of the resulting nanocrystal dispersions shows the disappearance of the characteristic C-H stretches from oleate, which is indicative of effective ligand removal (Figure S1). Zeta potential measurements confirm that long-range electrostatic repulsions originating from the metal oxide surface stabilize the nanocrystals ($\zeta$ = +38, +41, and +42 mV for FeO$_x$, Sn:In$_2$O$_3$, and F,Sn:In$_2$O$_3$, respectively, Figure S2). We use SAXS to probe the colloidal stability of these dispersions and find that they scatter as dilute and stable individual spheres and cubes (Figure 1d-f). We attribute the slight deviations in the low q scattering, relative to the simulated form factors, to weak interparticle interactions.[50] Suppressing aggregation in the initial dispersion, before adding depletion attractions, is crucial to avoid introducing uncontrolled structures that would ultimately be convoluted with longer length scale structures formed by depletion-induced assembly. In addition, the average nanocrystal sizes and size distributions were estimated by fitting the SAXS of the dispersions: the radii of FeO$_x$ and Sn:In$_2$O$_3$ spheres are 4.63 ± 0.16 nm and 5.34 ± 0.48 nm, respectively, while the half edge length of F,Sn:In$_2$O$_3$ cubes is 4.36 ± 0.45 nm. Because the nanocrystals are of similar and uniform size, the same depletant can be used to induce attractions with an approximately equal range relative to the size of the nanocrystals.

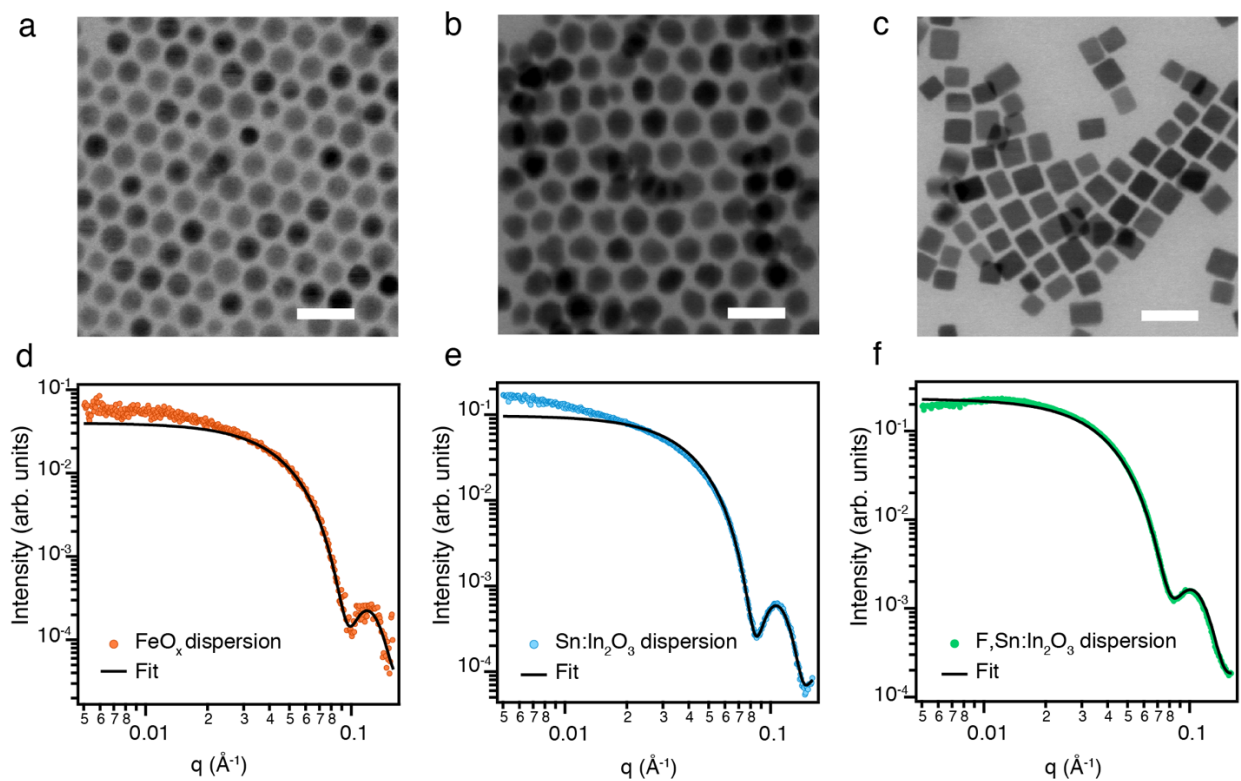

**Figure 1. Morphology, size, and colloidal stability of nanocrystals.** Scanning transmission electron microscopy (STEM) images of as-synthesized (a) $FeO_x$, (b) $Sn:In_2O_3$, (c) and $F,Sn:In_2O_3$ nanocrystals. Scale bars = 20 nm. SAXS of dilute (∼1 mg/mL) charge-stabilized nanocrystal dispersions in acetonitrile: (d) $FeO_x$ and spheroid model fit (R = 4.63 ± 0.16 nm), (e) $Sn:In_2O_3$ and spheroid model fit (R = 5.34 ± 0.48 nm), and (f) $F,Sn:In_2O_3$ and cuboid model fit (L = 4.36 ± 0.45 nm). R is the radius of a sphere and L is half the edge length of a cube.

Depletion gels were formed by adding PEG depletant (1 kDa number average molecular weight $M_n$, radius of gyration $R_g$ = 0.98 nm[44,51]) to dispersions in acetonitrile at a fixed nanocrystal volume fraction ($\phi_c$ = 0.04). We recently showed that PEG having this $M_n$ effectively drives the gelation of $Sn:In_2O_3$ nanocrystal spheres (R = 2.83 nm, $\phi_c$ = 0.04) dispersed in acetonitrile at a PEG concentration ([PEG]) of 587 mg/mL. In this study, the size ratio of depletant to nanocrystal is smaller by nearly a factor of 2, which shortens the effective length scale and lowers strength of the depletion attractions. Therefore, gelation occurs at higher [PEG]: 967 mg/mL for $FeO_x$ and $Sn:In_2O_3$ nanospheres (insets in Figure 2). $F,Sn:In_2O_3$ cubes gel at a lower [PEG] (351 mg/mL), which we attribute to their ability to pack face-to-face, leading to larger overlaps of excluded volumes and therefore stronger depletion attractions compared to the spheres. In each case, mixing the nanocrystals with lower [PEG] resulted in stable flowing dispersions (Table S1). Likewise, a solution of pure PEG in acetonitrile at the equivalent concentration used to induce gelation remained transparent and did not show signs of precipitation (Figure S3), suggesting that gelation arises due to PEG-mediated attractions between nanocrystals.

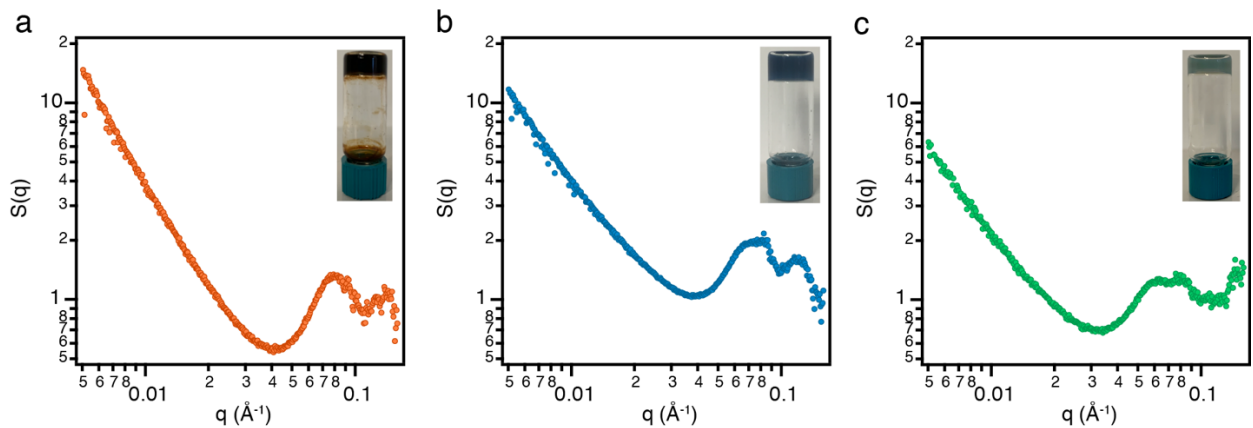

**Figure 2. Structure factor of depletion gels.** S(q) of nanocrystal gels with photographs of the corresponding vials (insets) for (a) $FeO_x$, (b) $Sn:In_2O_3$, and (c) $F,Sn:In_2O_3$ nanocrystals.

The gel structure was probed with SAXS to confirm the assembly of fully percolated networks (Figure 2). In particular, we isolate the structure factors S(q) of the gels by dividing the total scattering intensity by the form factors of the corresponding nanocrystals (Figure S4). S(q) diverges as q approaches zero for all samples and seems to follow a power law scaling characteristic of scattering from a mass fractal object.[52,53] To validate this observation and determine the number of structural length scales in the networks, we performed a derivative analysis (Figure S5). We find that each of the three gels scatters as a mass fractal with a single characteristic length scale. This single length scale suggests that the gel networks are formed as a result of depletion attractions between discrete nanocrystals, rather than depletion attractions between clusters of bridged nanocrystals as found previously for smaller $Sn:In_2O_3$ nanocrystals.[44] The fractal dimensions of the networks can be extracted from power law fits to S(q) in the region $q < 0.03$ Å$^{-1}$ (Figure S6). We obtained fractal dimensions of 2.0 for a $FeO_x$ nanocrystal gel and 1.7 for $Sn:In_2O_3$ and $F,Sn:In_2O_3$ nanocrystal gels, all of which fall within the range of fractal dimensions reported in the literature for percolated colloidal gels.[44,54-56] We also notice oscillatory features in S(q) at high q reminiscent of those that originate from hard sphere or sticky hard sphere interactions.[57-60] The primary peak appears at $q = 0.078$, $0.074$, and $0.062$ Å$^{-1}$, for $FeO_x$, $Sn:In_2O_3$, and $F,Sn:In_2O_3$ gels, respectively. These q values approximately correspond to the center-to-center distance of two adjacent nanocrystals (d= 8.05, 8.45, and 10.1 nm, respectively), which is consistent with the characteristic length scale for locally dense packing between nanocrystals in the gel.

To assess our proposed mechanism of depletion-driven gelation and better understand the influence of nanocrystal shape on the gelation threshold, we compare theoretical predictions for the phase behavior of the nanocrystal-depletant mixture to our experimental observations. Phase diagrams incorporating depletion attractions for both spherical[61] and cubic[62] nanocrystals are computed using free volume theory and scaled particle theory to determine the spinodal boundaries. Details of these calculations are found in the Supporting Information. At [PEG] above the spinodal curve, a homogeneous dispersed phase of nanocrystals is mechanically unstable with respect to fluctuations in nanocrystal concentration, and the dispersion undergoes phase separation via spinodal decomposition. Because of the strong interparticle depletion forces, colloids typically arrest in percolated networks,[35] so the spinodal boundary is the relevant one to compare with gelation in the experiments.[44] For both the sphere and the cube systems, we see good agreement between the predicted and experimental phase boundaries as a function of nanocrystal volume fraction ($\phi_c$) and [PEG] normalized by the polymer overlap concentration [PEG]$^*$ (Figure 3, Figure S7 for wider $\phi_c$ and [PEG] ranges, and Figure S8 for the FeO$_x$ nanocrystal spheres). Specifically, the nanocrystal gels lie at [PEG] above the spinodal, where gelation is expected, while the nanocrystal dispersions that remain flowing lie below the spinodal. Moreover, the phase diagrams clearly show that the gelation threshold is expected to occur at significantly lower [PEG] for the cubes compared to the spheres at all $\phi_c$ due to stronger depletion forces between flat surfaces compared to curved surfaces. This is confirmed in the experiments, where the Sn:In$_2$O$_3$ nanospheres remain flowing at [PEG]/[PEG]$^*$= 0.758 while the F,Sn:In$_2$O$_3$ nanocubes gel at the same [PEG].

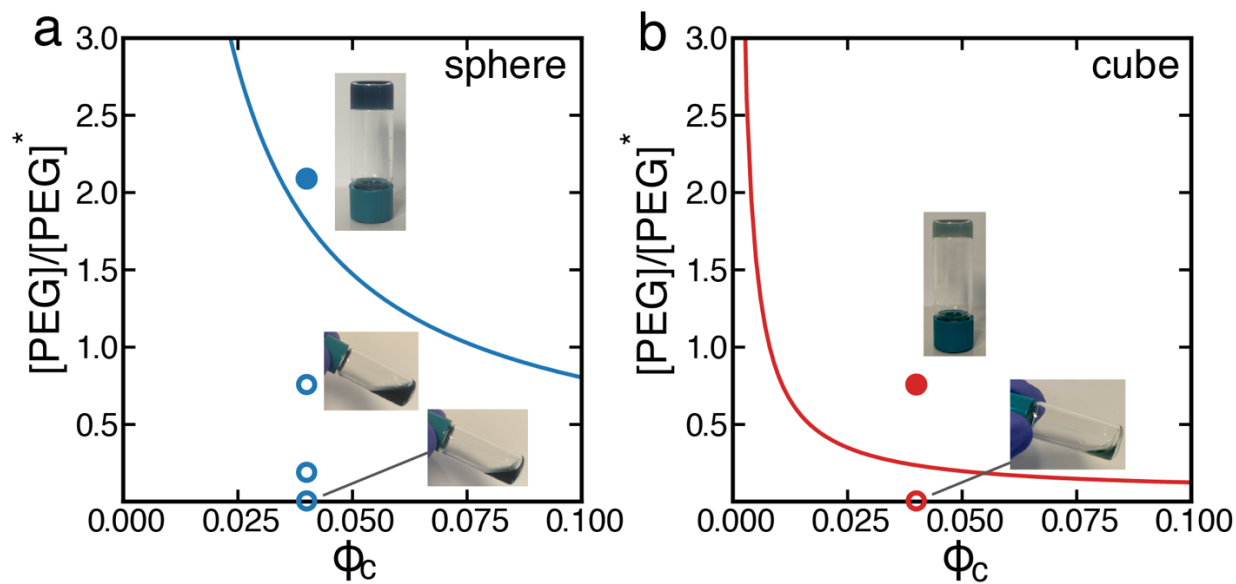

**Figure 3. Comparison of predicted phase behavior and experiments.** Theoretical spinodal boundaries (lines) and experimental observations (points) for (a) $Sn:In_2O_3$ nanocrystal spheres and (b) $F,Sn:In_2O_3$ nanocrystal cubes. The nanocrystal volume fraction is denoted by $\phi_C$, and the free polymer mass concentration [PEG] is normalized by the overlap concentration $[PEG]^*$. Details of this normalization are included in the Supporting Information. Open circles represent flowing dispersion samples while filled circles represent gel samples.

Depletion attractions were sufficient to account for the observed phase behavior of the nanocrystal dispersions, distinct from our previous work, where bridging of nanocrystals by PEG was observed to induce gelation at low [PEG]. For the nanocrystals studied here, monitoring the time evolution of S(q) for low [PEG] samples revealed only limited aggregation and no evidence for extended network formation, and the dispersions remained readily flowing even after several weeks (Figure S9-11). Although PEG adsorption on the metal oxides persists, PEG-mediated bridging attractions are significantly weaker, likely due to size constraints and smaller nanocrystal surface-to-volume ratios. In fact, we detect minimal clustering in all flowing dispersions. Aggregate sizes are only on the order of two adjacent nanocrystals (d = 19, 23, and 22 nm for $FeO_x$, $Sn:In_2O_3$, and $F,Sn:In_2O_3$, respectively) and do not grow significantly until after four weeks (Figure S12). We note that S(q) of the gels appears unchanged indicating the gel structures are not significantly changed by aging on this time scale.

In conclusion, we have demonstrated a gelation strategy broadly applicable to metal oxide nanocrystals. Nanocrystals of similar sizes, but different compositions and shapes, were assembled using the same PEG depletant and solvent combination. As expected theoretically, $FeO_x$ and $Sn:In_2O_3$ nanocrystal spheres formed gels at the same [PEG] while $F,Sn:In_2O_3$ nanocrystal cubes formed gels at a lower [PEG] due to the influence of shape on the strength of depletion attractions. Gel structure was characterized with SAXS, and our experimental observations were found to be consistent with depletion-driven gelation by comparison with the theoretically predicted phase behavior. Analysis of the SAXS data also supports the assembly of discrete nanocrystals rather than assembly of clusters of nanocrystals due to the excellent colloidal stability of our initial dispersions and the suppression of competing bridging attractions during the timeframe of gelation. Limiting nanocrystal aggregation apart from deliberate assembly driven by depletion attractions allows systematic control over the characteristic length scales of the microstructure in the gels. Finally, our findings motivate using depletion attractions to explore the connection between the gelation mechanism, the gel microstructure, and emergent properties. This work demonstrates the tunability and versatility of depletion attractions for nanocrystal assembly and provides a guide to expand the range of building blocks that can be used to incorporate targeted optical, electronic, or catalytic functionality in nanocrystal gels. Although varying nanocrystal composition and shape was our primary focus, within the paradigm described here, gels could be developed that utilize customized depletants among a wide variety of candidates (molecules, polymers, nanocrystals, etc), including ones that are stimuli responsive.[63-66]

## Associated Content

Details of nanocrystal synthesis, gel assembly, characterization (STEM, FTIR, zeta potential, SAXS), and theoretical methods.

## Author Information

Corresponding Authors:
Email: truskett@che.utexas.edu, milliron@che.utexas.edu## Notes

The authors declare no competing financial interest.

**Acknowledgements**

This research was primarily supported by the National Science Foundation (NSF) through the Center for Dynamics and Control of Materials: an NSF MRSEC under Cooperative Agreement No. DMR-1720595. SAXS data was collected at UT Austin with an instrument acquired under NSF MRI grant (CBET-1624659). Additional support was provided by NSF (CHE-1905263, CBET-1704634), the Fulbright Program (Grant No. IIE-15151071 to SHC), and the Welch Foundation (F-1848 and F-1696).

**Table of Contents Graphic**

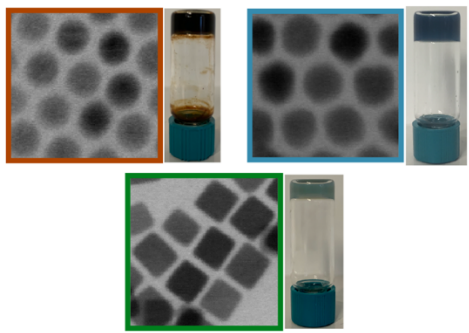